\documentclass[sigplan,10pt]{acmart}

\acmJournal{PACMPL}
\acmVolume{1}
\acmNumber{} 
\acmArticle{1}
\acmYear{2020}
\acmMonth{1}
\acmDOI{} 
\startPage{1}

\setcopyright{none}

\usepackage{booktabs}   
\usepackage{subcaption} 
\usepackage{multicol}
\usepackage{soul}
\usepackage{todonotes}

\usepackage{cedilleverbatim}

\usepackage{proof}
\usepackage{alltt}

\newtheorem{requirement}{Requirement}

\newcommand{\abs}[4]{{#1}\, \mathit{#2}\! : \! #3.\, #4}
\newcommand{\absu}[3]{{#1}\, \mathit{#2}.\, #3}
\mathchardef\mhyph="2D 
\newcommand{\ann}[2]{\mathit{#1}\! : \! #2}

\newcommand{\vars}[1]{{\overline{#1}}}
\newcommand{\elab}{\hookrightarrow}
\newcommand{\reduce}{\ensuremath{\rightsquigarrow}}
\newcommand{\piforall}{^{\Pi}_{\forall}}
\newcommand{\lamLam}{^{λ}_{Λ}}
\newcommand{\indsche}[3]{\ensuremath{\text{Ind}[#1,#2,#3]}}
\newcommand{\indel}[5]{\ensuremath{\text{IndEl[}#1,#2,#3,#4,#5\text{]}}}

\newcommand{\les}{\ensuremath{\leqslant}}

\newcommand{\wf}{\ensuremath{w\!f}}
\newcommand{\mufix}[4]{\ensuremath{μ\ \mathit{#1}.\ \mathit{#2}\ @#3\ \{\ #4\}}}
\newcommand{\mufixu}[3]{\ensuremath{μ\ \mathit{#1}.\ \mathit{#2}\ \{\ #3\}}}
\newcommand{\mumat}[4]{\ensuremath{μ'\texttt{<$\mathit{#1}$>}\ #2\ @#3\ \{\ #4\}}}
\newcommand{\mumatu}[2]{\ensuremath{μ'\ #1\ \{\ #2\}}}
\newcommand{\mup}{\ensuremath{μ'}}

\newcommand{\splab}[1]{\ensuremath{^{\text{#1}}}}
\newcommand{\elalab}[1]{\ensuremath{\overset{\text{#1}}{\elab}}}
\newcommand{\elapos}{\ensuremath{\overset{\text{+}}{\elab}}}
\newcommand{\elales}{\ensuremath{\elab}}

\newcommand{\genid}[2]{\ensuremath{\texttt{#1}\!\mathit{#2}}}

\DeclareUnicodeCharacter{0B7}{\ensuremath{\cdot}}

\begin{document}

\title[]{Elaborating Inductive Definitions and Course-of-Values Induction
  in Cedille}


\author{Christopher Jenkins}
\orcid{0000-0002-5434-5018}             
\affiliation{
  \position{}
  \department{Computer Science}              
  \institution{University of Iowa}            
  \streetaddress{}
  \city{Iowa City}
  \state{Iowa}
  \postcode{52240}
  \country{United States of America}                    
}
\email{christopher-jenkins@uiowa.edu}          

\author{Colin McDonald}
\orcid{nnnn-nnnn-nnnn-nnnn}             
\affiliation{
  \position{}
  \department{Computer Science}             
  \institution{University of Iowa}           
  \streetaddress{}
  \city{Iowa City}
  \state{Iowa}
  \postcode{52240}
  \country{United States}                   
}
\email{colinmcd0731@gmail.com}         

\author{Aaron Stump}
\orcid{nnnn-nnnn-nnnn-nnnn}             
\affiliation{
  \position{Professor}
  \department{Computer Science}             
  \institution{University of Iowa}           
  \streetaddress{}
  \city{Iowa City}
  \state{Iowa}
  \postcode{52240}
  \country{United States}                   
}
\email{aaron-stump@uiowa.edu}         


\begin{abstract}
  In the Calculus of Dependent Lambda Eliminations (CDLE), a pure Curry-style
  type theory, it is possible to generically λ-encode inductive datatypes which
  support \textit{course-of-values} (CoV) induction. We present a datatype
  subsystem for Cedille (an implementation of CDLE) that provides this feature
  to programmers through convenient notation for declaring datatypes and for
  defining functions over them by case analysis and fixpoint-style recursion
  guarded by a type-based termination checker. We demonstrate that this does not
  require extending CDLE by showing how datatypes and functions
  over them elaborate to λ-encodings, and proving that this elaboration is type-
  and value-preserving. This datatype subsystem and elaborator are
  implemented in Cedille, establishing for the first time a complete translation
  of inductive definitions to a small pure typed λ-calculus.
  
\end{abstract}



\maketitle

\section{Introduction}
\label{sec:intro}
Algebraic datatypes (ADTs) are a popular feature of functional programming
languages that combine a concise scheme for declaring datatypes and
their constructors with an intuitive mechanism for defining functions over them
by pattern matching and recursion.
Their popularity extends to implementations of type theories, wherein
properties of data are proven using the same mechanisms as for defining
functions over them.
However, a wrinkle in bringing ADTs to proof assistants based on type theory
is concern for the \emph{de Bruijn criterion}
~\cite{Geu09_Proof-Assistants-History}, i.e., that the assistant produce
proof objects checkable by an implementation of a small kernel theory.
Of particular concern are \emph{termination checking} for recursive definitions
and \emph{positivity checking} for data declarations, since to 
maintain logical soundness implementations must usually ensure functions and
proofs are well-founded (\cite{Me91_Inductive-Types-20-Lambda-Calculus} showed
that non-positive datatypes can be used to define looping terms, without any
apparent recursion in the language).

The most common approaches to both positivity and termination checking are
\emph{syntactic}: for the former, this involves tracking, in the types of
constructor arguments, the number of arrows of which a recursive occurrence of a
datatype is to the left (c.f. \cite[][Section 4.5.2]{Coq17_Manual}); for the
latter, recursive invocations are allowed only on subdata revealed by pattern
matching within the function~\cite{Gi95_Guarded-Defs-Recursion}.
Any such syntactic criteria must usually be implemented in the kernel language also,
increasing its complexity.
This situation is especially unfortunate for termination checking, as
simpler syntactic guards are brittle so it is tempting to make these more
sophisticated to grow the set of accepted definitions.

Happily, positivity and termination checking have more \emph{semantic} approaches:
for the former, explicit evidence of positivity can be required to form an
inductive type \cite{Mat02_Tarski-and-Monotone-Inductive-Types}, or polarity
annotations can be added to the language of kinds
\cite{AGHH06_Polarized-Subtyping-for-Sized-Types}; for the latter,
\emph{type-based} termination checking augments the type system itself with some
notion of the size of datatypes \cite{Ab10_Sized-Types,BFGPU04_Type-Based-Termination}.
Such principled extensions have more modest impact on the complexity of the
kernel language, though can require reworking of pre-existing meta-theoretic results.

In pure type systems, datatypes are defined with λ-encodings that
combine case analysis and recursion into a single scheme that ensures
termination.
Cedille
~\cite{St17_CDLE,St18_Cedille-Syntax-Semantics} is a dependently typed
programming language which overcomes some traditional shortcomings of
λ-encodings in type theory (e.g., underivability of
induction~\cite{Ge01_No-Induction-2O-DTT}).
Cedille's core theory, the
\textit{Calculus of Dependent Lambda Eliminations} (CDLE), is a compact pure
Curry-style type theory with no primitive notion of inductive datatypes.
Instead, and as shown by \cite{FBS18_Efficient-Mendler}, it is possible to
generically derive the induction principle for λ-encoded data using a
Mendler-style of encoding that features constant-time predecessors and a
linear-space representation. Furthermore, \cite{FDJS18_CoV-Ind} show how to
further augment this with \textit{course-of-values} (\textit{CoV}) induction, an
expressive scheme wherein recursive calls are allowed on nested subdata at
unbounded depth and whose well-foundedness is tricky to convey to
syntactic termination checkers.

\paragraph{\textbf{Contributions}}
Most programmers (and type theorists!) do not wish to work directly with λ-encodings.
Building off the work of \cite{FBS18_Efficient-Mendler, FDJS18_CoV-Ind}, in this
paper we add \emph{language-level} support for inductive types in Cedille
by presenting a datatype subsystem with convenient notation for declaring
datatypes and functions defined over them using pattern matching and
fixpoint-style recursion. In particular, we:
\begin{itemize}
\item design a \textit{semantic} (type-based) termination checker based on
  \textit{CoV pattern matching}, a novel feature allowing
  Cedille to accept recursive definitions expressed as CoV induction schemes
  (Section \ref{sec:cov-induction});
\item show how datatype declarations and functions over them are elaborated to
  λ-encodings in Cedille (Sections \ref{sec:elab-data} and
  \ref{sec:datatype-functions}); and
\item prove that elaboration is
  type- and value-preserving, demonstrating that the above can be achieved
  without extension of CDLE (Sections
  \ref{ssec:elab-datatypes}, \ref{ssec:elab-mu}, and \ref{ssec:mu-reduce}).
\end{itemize}
The datatype system and elaborator are implemented in Cedille
(\url{github.com/cedille/cedille}). Our approach demonstrates that inductive
definitions in constructive type theory can be soundly translated down to a very
small pure type system. Indeed, there is already a translation from Cedille
1.0.0 (which does not have datatypes) to Cedille Core, a minimal specification
of CDLE implemented in $\sim$1K Haskell LoC. This paper and its proof
appendix treats formally only the elaboration of non-indexed datatypes to
Cedille 1.0.0.

The remainder of this paper is organized as follows: in Section \ref{ssec:cdle}
we review CDLE; in Section \ref{sec:cedille-datatypes} we describe
datatype system using standard examples; Section \ref{sec:cov-induction}
explains CoV pattern matching; in Section \ref{sec:generic-interface} we
describe the elaborator interface; in Section \ref{sec:elab-data} we formally treat
elaboration of datatype declarations; in Section \ref{sec:datatype-functions}
we explain elaboration of functions over data using CoV pattern-matching and
recursion; and in Sections \ref{sec:related-future-work} and
\ref{sec:conclusion} we discuss related and future work.

\subsection{Background: CDLE}
\label{ssec:cdle}
\begin{figure*}
  \begin{subfigure}{0.75\linewidth}
    \[\small
      \begin{array}{c}
        \begin{array}{cc}
          \infer{
          \Gamma \vdash \beta : \{t ≃ t\}
          }{
          \quad \mathit{FV}(t) \subseteq \mathit{dom}(\Gamma)
          }
          &
            \infer{
            \Gamma \vdash \varphi\ t -\ t_1\ \{t_2\} : T
            }{
            \Gamma \vdash t : \{t_1 ≃ t_2 \}
            \quad \Gamma \vdash t_1 : T
            }
        \end{array}
        \\ \\
        \begin{array}{ccc}
        \infer{
          \Gamma\vdash [t_1,t_2] : \abs{\iota}{x}{T_1}{T_2}
          }{
          \Gamma\vdash t_1 : T_1
          \quad \Gamma\vdash t_2 : [t_1/x]T_2
          \quad |t_1| = |t_2|
          }
        &
          \infer{
          \Gamma\vdash t.1 : T_1
          }{
          \Gamma\vdash t : \abs{\iota}{x}{T_1}{T_2}
          }
        &
          \infer{
          \Gamma\vdash t.2 : [t.1/x]T_2
          }{
          \Gamma\vdash t : \abs{\iota}{x}{T_1}{T_2}
          }
        \end{array}
        \\ \\
        \begin{array}{cc}
          \infer{
          \Gamma\vdash\abs{\Lambda}{x}{T}{t'} : \abs{\forall}{x}{T}{T'}
          }{
          \Gamma,\ann{x}{T}\vdash t':T'
          \quad x\not\in\textit{FV}(|t'|)
          }
          &
            \infer{
            \Gamma\vdash t\ \mhyph t' : [t'/x]T
            }{
            \Gamma\vdash t : \abs{\forall}{x}{T'}{T}
            \quad \Gamma\vdash t' : T'
            }
        \end{array}
      \end{array}
    \]
  \end{subfigure}%
  \begin{subfigure}{0.2\linewidth}
    \[ \small
      \begin{array}{rcl}
        |\beta| & = & \absu{λ}{x}{x}
        \\ |\varphi\ t - t_1\ \{t_2\}| & = & |t_2|
        \\ |[ t_1 , t_2 ]| & = & |t_1|
        \\ |t.1| & = & |t|
        \\ |t.2| & = & |t|
        \\ |\abs{\Lambda}{x}{T}{t}| & = & |t|
        \\ |t\ \mhyph t'| & = & |t|
      \end{array}
    \]
  \end{subfigure}
  \caption{Kinding, typing, and erasure for a fragment of CDLE}
  \label{fig:cdle-type-constructs}
\end{figure*}
We review the Calculus of Dependent Lambda Eliminations (CDLE), the type
theory of Cedille. CDLE is an extension of the impredicative Curry-style (i.e.,
extrinsically typed) Calculus of Constructions (CC) that adds three new type
constructs: equality of untyped terms ($\{t ≃ t'\}$); dependent intersections
($\abs{ι}{x}{T}{T'}$) of \cite{Ko03_Dependent-Intersection}; and the
implicit (erased) products ($\abs{∀}{x}{T}{T'}$) of \cite{Mi01_ICC}. The
pure term language of CDLE is the untyped $\lambda$-calculus; to
make type checking algorithmic, terms in Cedille are type annotated, and
definitional equality of terms is modulo erasure of annotations.
The typing and erasure rules for the fragment of CDLE relevant to this paper are
given in Figure \ref{fig:cdle-type-constructs}, with a full listing given in
\cite{St18_Cedille-Syntax-Semantics} and this paper's proof appendix.

\paragraph{Equality}
$\{t_1 \simeq t_2\}$ is the type of proofs that the erasures of $t_1$ and
$t_2$ (resp. $|t_1|$ and $|t_2|$) are equal. It is introduced with
$\beta$ (erasing to $\absu{λ}{x}{x}$) proving $\{t
\simeq t\}$ for any untyped term $t$. Combined with definitional equality,
$\beta$ can prove $\{t_1 ≃ t_2\}$ for any $\beta\eta$-convertible
$t_1$ and $t_2$ whose free variables are declared in the context.
Equality proofs can be eliminated with $\varphi$, where the expression $\varphi\
t - t_1\ \{t_2\}$ (erasing to $|t_2|$) casts $t_2$ to the 
  type of $t_1$  when $t$ proves $t_1$ and $t_2$ are equal. 

\paragraph{Dependent intersection}
$\abs{\iota}{x}{T}{T'}$ is the type of terms $t$ which can be assigned
  both type $T$ and $[t/x]T'$, and in the annotated language is introduced by
$[t_1 , t_2]$, where $t_1$ has type $T$, $t_2$ has type $[t_1/x]T'$, and $|t_1|
=_{\beta\eta} |t_2|$. Dependent intersections are eliminated with
projections $t.1$ and $t.2$, selecting resp. the view that term $t$ has type $T$
or $[t.1/x]T'$

\paragraph{Implicit product}
$\abs{\forall}{x}{T}{T'}$ is the type of dependent functions with an
erased argument of type $T$ and a result of type $T'$. They are introduced
with $\abs{\Lambda}{x}{T}{t}$, provided $x$ does not occur free in $|t|$, and they
are eliminated with erased application $t_1\ \mhyph t_2$. Erased arguments play no
computational role and exist solely for the purposes of typing.

Figure \ref{fig:cdle-type-constructs} omits typing and erasure rules for the
term and type constructs of CC. In terms, all type annotations and abstractions
(also using Λ) are erased, and the argument of term to type applications (written $t ·S$) is erased.
In types, ∀ and λ resp. quantify and abstract over types, and type to type
application is written $T ·S$.
In code listings, we omit type arguments and annotations when Cedille can infer these.

\subsection{Datatypes in Cedille}
\label{sec:cedille-datatypes}
\begin{figure*}[h]
  \small
  \begin{subfigure}{0.5\linewidth}
    \caption{Datatype declarations}
    \label{sfig:data-decl}
\begin{alltt}
data Bool: ★ = tt: Bool  | ff: Bool.
data Nat:  ★ = zero: Nat | suc: Nat ➔ Nat.
data List (A: ★): ★
= nil: List | cons: A ➔ List ➔ List.
\end{alltt}
  \end{subfigure}%
  \begin{subfigure}{0.5\linewidth}
    \caption{Functions}
    \label{sfig:fun}
\begin{alltt}
pred: Nat ➔ Nat = λ n. μ' n \{zero ➔ n | suc n' ➔ n'\}.
add: Nat ➔ Nat ➔ Nat
= λ m. λ n. μ addN. m \{zero ➔ n | suc m' ➔ suc (addN m')\}.

\end{alltt}
  \end{subfigure}
  \caption{Example datatype declarations and functions}
  \label{fig:cedille-data-standard}
\end{figure*}%
\paragraph{\textbf{Declarations}}%
Figure \ref{sfig:data-decl} show definitions of well-known types using Cedille's
datatype subsystem.
The general
scheme for declaring datatypes in Cedille should be straightforward to anyone familiar with
GADTs in Haskell or with dependently typed languages like Agda, Coq, or Idris.
We note some differences from the usual convention below.

\begin{itemize}
\item Occurrences of the inductive type being defined are not written applied to
  its parameters.
  For example, the constructor \texttt{nil} is written having type
  \texttt{List} rather than \(\texttt{List}\ ·A\); used outside of the datatype
  declaration, \texttt{nil} has the usual type \(\abs{∀}{A}{★}{\texttt{List}\
    ·A}\).
\item In constructor types, recursive occurrences of the datatype (such as
  \texttt{\underline{Nat}} in \(\texttt{suc}: \texttt{\underline{Nat}} ➔
  \texttt{Nat}\) 
  must be positive, but \textit{need not be} strictly
  positive (\cite{Bl05_Inductive-Types-CAC} showed strict positivity is
    not needed for small datatypes).
\item Declarations can only refer to the datatype itself and prior
  definitions. Inductive-recursive and inductive-inductive definitions are not
  part of this proposal.
\end{itemize}

\paragraph{\textbf{Functions}}
To continue to familiarize the reader with Cedille's syntax, Figure
\ref{sfig:fun} shows a few standard examples of functional and dependently typed
programs.
Function \texttt{pred} introduces
operator {\mup} for \textit{CoV pattern matching}, where it is used for standard
pattern matching. Its operational semantics (Section \ref{ssec:mu-reduce}) is
the usual case branch selection. In \texttt{pred}, {\mup} is given scrutinee $n$ of type
\texttt{Nat} and a case tree with branches for each constructor of \texttt{Nat}.

Function \texttt{add} introduces operator μ for \textit{CoV
  induction} by combined pattern matching and recursion; the distinction between
pattern matching by μ and {\mup} is made clear in Section \ref{sec:cov-induction}.
Its operational semantics is combined case branch selection and fixpoint
unrolling. Here, μ is used for standard structurally recursive definitions: in
\texttt{add} it is used to define function $\mathit{addN}$ (so named
because it adds $n$ to its argument), and in the successor case $\mathit{addN}$
is recursively invoked on the subdata $m'$ revealed by the constructor pattern
\(\texttt{suc}\ m'\).


\section{Course-of-Values Recursion}
\label{sec:cov-induction}
This section explains \textit{course-of-values (CoV) pattern matching}, a 
feature that is the basis of Cedille's type-based termination checker. The
example of division given in this section is similar to one appearing in
\cite{FDJS18_CoV-Ind}; whereas they use their generic development as a library
to implement this, we use their development as a back-end, and the example
illustrates CoV recursion in our surface language. Due to space restrictions we
do not discuss here an example of CoV \emph{induction} in the surface language,
though this too is supported by the datatype subsystem (see Sections
\ref{sec:generic-interface} and \ref{sec:datatype-functions}).

\paragraph{\textbf{Termination checking}}
In general purpose functional languages, programmers are free to define functions
using powerful recursion schemes, including general recursion.
Users of implementations of type theories are usually not afforded such freedom, as
these implementations must usually ensure recursive definitions are well-founded or
risk logical unsoundness.
To that end, it is common to use a termination checker
implementing a \textit{syntactic guard}, enforcing that recursive calls are made
only on terms revealed by case analysis on arguments of the function.

Unfortunately, syntactic termination checkers are usually unable to determine
that complex recursion schemes are well-founded
(\cite{BKS16_Partiality-and-Recurstion-in-ITP,BFGPU04_Type-Based-Termination}).
Consider an intuitive definition of division by iterated
subtraction. In a Haskell-like language, programmers write: {\small
\begin{alltt}
zero    / d = 0
(suc n) / d = if (suc n < d)
              then zero else suc ((n - (d - 1)) / d)
\end{alltt}
}\noindent
This definition is guaranteed to terminate for all inputs, as the first
argument to the recursive call, $n\ \mhyph\ (d\ \mhyph\ 1)$, is smaller than the
original argument $\texttt{suc}\ n$. As innocuous as this definition may seem to
functional programmers, it poses a difficulty for syntactic termination
checkers, as $n\ \mhyph\ (d \mhyph 1)$ is not an expression produced by
case analysis of $n$ within the definition of division but an \textit{arbitrary}
predecessor produced by $d-1$ iterations of case analysis. This is the
\emph{course-of-values} recursion scheme (categorically, \textit{histomorphism});
it is guaranteed to be terminating, but this fact is difficult to communicate to
syntactic termination checkers!

\subsection{Course-of-values Pattern Matching}
Cedille implements \emph{type-based} termination checking that is powerful
enough to accept functions defined using the CoV recursion scheme.
At its heart
is a feature we call \textit{CoV pattern matching}, invoked by μ', which can be
used to define a version of division written close to the intuitive way, only
requiring some typing annotations to guarantee termination.

Termination checking in Cedille works by replacing, in the types of subdata in pattern
guards of inductive μ-expressions (but not \mup), the recursive occurrences of a
datatype with an abstract (as in, universally quantified) type. This abstract
type and not the usual datatype is the type of legal arguments for recursive
calls. Crucially, CoV pattern matching with {\mup} \textit{preserves} this type
in the subdata revealed by case patterns, meaning users can write versions of
e.g. predecessor and subtraction which can be used to compute values which are then given to
recursive calls of division; furthermore, they are easily \textit{reused for
  ordinary numbers}. Figure
\ref{fig:cov-divide} gives these and other auxiliary definitions.

\begin{figure*}
\small
\begin{alltt}
predCoV: ∀ N: ★. ∀ is: Is/Nat ·N. N ➔ N = Λ N. Λ is. λ n. μ'<is> n \{zero ➔ n | suc n' ➔ n'\}.

minusCoV: ∀ N: ★. ∀ is: Is/Nat ·N. N ➔ Nat ➔ N
= Λ N. Λ is. λ m. λ n. μ mMinus. n \{
  | zero ➔ m
  | suc n' ➔ predCoV -is (mMinus n') \}.
minus = minusCoV -is/Nat.

lt: Nat ➔ Nat ➔ Bool = λ m. λ n. μ' (minus (suc m) n) \{zero ➔ tt | suc r ➔ ff\}.
ite: ∀ X: ★. Bool ➔ X ➔ X ➔ X = Λ X. λ b. λ t. λ f. μ' b \{tt ➔ t | ff ➔ f\}.

divide: Nat ➔ Nat ➔ Nat = λ n. λ d. μ divD. n \{
| zero ➔ zero
| suc pn ➔ [pn' = to/Nat -isType/divD pn] - [diff = minusCoV -isType/divD pn (pred d)] -
  ite (lt (suc pn') d) zero (suc (divD diff)) \}.
\end{alltt}
  \caption{Division using course-of-values recursion}
  \label{fig:cov-divide}
\end{figure*}

\paragraph{\textbf{Global declarations}}
We first explain the types and definitions of \texttt{predCoV} and
\texttt{minusCoV}. In \texttt{predCoV} we see the first use of predicate
\texttt{Is/Nat}. Every datatype declaration in Cedille additionally
introduces three global names derived from the datatype's name. For
\texttt{Nat}, these are:

\begin{itemize}
\item \(\texttt{Is/Nat}: ★ ➔ ★\)

  A term of type \(\texttt{Is/Nat} ·N\) is a witness that any term of type
  \(N\) may be treated as if it has type \texttt{Nat} for CoV pattern matching.

\item \(\texttt{is/Nat} : \texttt{Is/Nat} · \texttt{Nat}\) is the trivial
\texttt{Is/Nat} witness.
\item \(\texttt{to/Nat}: \abs{∀}{N}{★}{\abs{∀}{is}{\texttt{Is/Nat} ·N}{N ➔
      \texttt{Nat}}}\)

  \texttt{to/Nat} is a function that coerces a term of type $N$ to
  \texttt{Nat}, given a witness $\mathit{is}$ that $N$ ``is'' \texttt{Nat}.
\end{itemize}


In \texttt{predCoV} the witness $\mathit{is}$ of type \(\texttt{Is/Nat} ·N\) is
given explicitly to {\mup} with the notation \(\mup\texttt{<}is\texttt{>}\),
allowing argument $n$ (of type $N$) to be a legal scrutinee for \texttt{Nat}
pattern matching. Reasoning parametrically, the only ways \texttt{predCoV} can
produce an $N$ output (i.e, preserve the abstract type of its argument) are by returning $n$
itself or some subdata produced by CoV pattern matching on it -- the predecessor
$n'$ also has type $N$. Thus, the type signature of \texttt{predCoV} has the
following intuitive reading: it produces a number no larger than its argument,
since a result like \(\texttt{suc}\ (\texttt{to/Nat}\ \mhyph \mathit{is}\ n)\)
would be type-incorrect. Note though that this reading is informal and outside of
the theory, whereas approaches based on sized types use explicit size indices to
track structural decrease.

\paragraph{\textbf{Code Reuse}}
The reader may now wonder what the relation is between \texttt{predCoV}
and the earlier \texttt{pred} of Figure \ref{sfig:fun}. The
μ'-expression of \texttt{pred} with the witness given explicitly is:
{\small
\begin{alltt}
μ'<is/Nat> n \{zero ➔ n | suc n' ➔ n'\}
\end{alltt}
}
\noindent In \texttt{pred}, the global witness \texttt{is/Nat} of type
\(\texttt{Is/Nat} ·\texttt{Nat}\) need not be passed explicitly, as it is
inferable by the type
\texttt{Nat} of the scrutinee $n$. Furthermore, \texttt{pred} and
\texttt{predCoV} are definitionally equal, as these witnesses are erased
from μ'-expressions (below \_ indicates an anonymous proof): {\small
\begin{alltt}
_ : \{pred ≃ predCoV\} = β.
\end{alltt}
}

This leads to a style of programming where, when possible, functions are defined
over an abstract type $N$ for which e.g. \(\texttt{Is/Nat} ·N\) holds, and the
usual versions of functions \textit{reuse} these as a special case. This is how
\texttt{minus} is defined -- by specializing \texttt{minusCoV}
with the trivial witness \texttt{is/Nat}.

The type signature of \texttt{minusCoV} similarly yields a reading that its
result is no larger than its first argument. In the successor case,
\texttt{predCoV} is given the (erased) witness $\mathit{is}$. That
\texttt{minusCoV} preserves the type of its argument after $n$ uses of
\texttt{predCoV} is precisely what allows it to appear in an 
argument to recursive functions over \texttt{Nat}.
Function \texttt{minus} is used to define
\texttt{lt}, the Boolean predicate deciding whether its first argument is less
than its second; \texttt{ite} is the usual definition of a conditional
expression by case analysis on \texttt{Bool}.

\paragraph{Division}
The last definition, \texttt{divide}, is as expected except for the successor
case.
Here, we make let bindings for $\mathit{pn'}$ and $\mathit{diff}$, the syntax
for which in Cedille is \([x = t] - t'\) analogous to \(\texttt{let}\ x = t\
\texttt{in}\ t'\).
Term $\mathit{pn'}$ is the coercion to \texttt{Nat} of the predecessor of the
dividend $\mathit{pn}$, using the
as-yet unexplained \texttt{Is/Nat} witness \texttt{isType/divD}.
Term $\mathit{diff}$ is the difference (computed using \texttt{minusCoV}) between $\mathit{pn}$ and
\(\texttt{pred}\ d\).
Note that $\mathit{diff}$ is guaranteed to be smaller than
the original pattern \(\texttt{suc}\ pn\). Finally, we test whether the dividend
is less than the divisor: if so, return \texttt{zero}; if not, divide
$\mathit{diff}$ by $d$ and increment.
The only parts of \texttt{divide}
requiring further explanation are the witness \texttt{isType/divD} and
the type of $\mathit{pn}$, which are the keys to CoV recursion in Cedille.

\paragraph{\textbf{Local declarations}}
Within the body of the μ-expression defining recursive function
\texttt{divD} over scrutinee $n$ of type \texttt{Nat}, the following
names are automatically bound:
\begin{itemize}
\item \(\texttt{Type/divD}: ★\), the type of recursive occurrences of \texttt{Nat}
  in the types of variables bound in constructor patterns (such as $pn$).
\item \(\texttt{isType/divD}: \texttt{Is/Nat} ·\texttt{Type/divD}\), a witness
  that terms of type \texttt{Type/divD} may used for CoV pattern matching.
\item \(\texttt{divD}: \texttt{Type/divD} ➔ \texttt{Nat}\),
  the recursive function being defined, accepting only terms of the
  abstract type \texttt{Type/divD}. This restriction guarantees that
  \texttt{divD} is only called on expressions smaller than the previous
  argument to recursion.
\end{itemize}

The reader is now invited to revisit the definitions of Figure
\ref{fig:cedille-data-standard}, keeping in mind that the μ-expression of
\texttt{add}, for example, the subdata $m'$ 
in pattern guard \(\texttt{suc}\ m'\)
has an abstract type, and the recursively defined \texttt{addN} only accepts
arguments of such a type.
With this understood, so to is 
\texttt{divide}: predecessor $\mathit{pn}$ has type \texttt{Type/divD}, witness
\texttt{isType/divD} has type \(\texttt{Is/Nat} · \texttt{Type/divD}\) and so the local
variable $\mathit{diff}$ has type \texttt{Type/divD} as required by
\texttt{divD}.

\section{Elaboration Interface}
\label{sec:generic-interface}%
\begin{figure*}[h]
  \begin{subfigure}{1\linewidth}
    \caption{Casts, positivity, and type fixpoints}
    \label{sfig:elab-toolkit}
    \[
      \begin{array}{c}
      \begin{array}{ccc}
        \infer{
         \Gamma \vdash \texttt{Cast} ·A ·B : ★
        }{
         \Gamma \vdash A : ★ \quad \Gamma \vdash B : ★
        } 
        &
          \infer{
           \Gamma \vdash \texttt{intrCast}\ f\ p : \texttt{Cast} ·A ·B
          }{
           \Gamma \vdash f : A ➔ B \quad \Gamma \vdash p : \abs{Π}{x}{A}{\{f\ x
          ≃ x\}}
          }
        &
          \infer{
           \Gamma \vdash \texttt{elimCast}\ \mhyph c \cong \absu{λ}{x}{x} : A ➔ B
          }{
           \Gamma \vdash c : \texttt{Cast} ·A ·B
          }
      \end{array}
      \\ \\
      \begin{array}{cc}
        \infer{
         \Gamma \vdash \texttt{Mono} ·F : ★
        }{
         \Gamma \vdash F : ★ ➔ ★
        }
        &
          \infer{
           \Gamma \vdash \texttt{intrMono}\ f : \texttt{Mono} ·F
          }{
           \Gamma \vdash f : \abs{∀}{X}{★}{\abs{∀}{Y}{★}{\texttt{Cast} ·X ·Y ➔ \texttt{Cast}
          ·(F ·X) ·(F ·Y)}}
          }
      \end{array}
        \\ \\
        \begin{array}{cc}
          \infer{
           \Gamma \vdash \texttt{elimMono}\ \mhyph \mathit{im}\ \mhyph c \cong \absu{λ}{x}{x}
          : F ·A ➔ F ·B
          }{
          \Gamma \vdash \mathit{im} : \texttt{Mono} ·F
          \quad \Gamma \vdash c : \texttt{Cast} ·A ·B
          }
        & 
          \infer{
        \Gamma \vdash \texttt{Fix} ·F\ \mathit{im} : ★
        }{
        \Gamma \vdash F : ★ ➔ ★
        \quad \Gamma \vdash \mathit{im} : \texttt{Mono} ·F
          }
          \end{array}
        \\ \\
        \begin{array}{cc}
          \infer{
           \Gamma \vdash \texttt{in}\ \mathit{xs} : \texttt{Fix} ·F\ \mathit{im}
          }{
          \Gamma \vdash F : ★ ➔ ★
          \quad \Gamma \vdash \mathit{im} : \texttt{Mono} ·F
          \quad \Gamma \vdash \mathit{xs} : F ·(\texttt{Fix} ·F\ \mathit{im})
          }
          &
            \infer{
            \Gamma \vdash \texttt{out}\ \mathit{x} : F ·(\texttt{Fix} ·F\ \mathit{im})
            }{
            \Gamma \vdash F : ★ ➔ ★
            \quad \Gamma \vdash \mathit{im} : \texttt{Mono} ·F
            \quad \Gamma \vdash \mathit{x} : \texttt{Fix} ·F\ \mathit{im}
            }
        \end{array}
      \end{array}
    \]
  \end{subfigure}
  \begin{subfigure}{1\linewidth}
    \caption{Generic induction principle for λ-encoded data}
    \label{sfig:elab-mendler}
    {\small
\begin{alltt}
module GenericCoV (F: ★ ➔ ★) \{im: Mono ·F\}

D: ★ = Fix ·F im.

PrfAlg: (D ➔ ★) ➔ ★ = λ P: D ➔ ★. ∀ R: ★. ∀ c: Cast ·R ·D. Π o: R ➔ F ·R. ∀ oeq: \{o ≃ out\}.
    (Π x: R. P (elimCast -c x)) ➔ Π xs: F ·R. P (in (elimMono -im -c xs)).

induction: ∀ P: D ➔ ★. PrfAlg ·P ➔ Π x: D. P x = <..>
\end{alltt}
}
  \end{subfigure}
  \caption{Generic library}
  \label{fig:elab-generic}
\end{figure*}
The generic library of \cite{FBS18_Efficient-Mendler,FDJS18_CoV-Ind} derives
inductive datatypes using \textit{Mendler-style F-algebras}, so we begin with a
brief description of these. For a more thorough treatment of the expressive
power of Mendler-style algebras, see \cite{AS11_Mendler-Hierarchy}.

\paragraph{\textbf{Mendler-style \(F\)-algebras}} It is well understood that an
inductive datatype $D$ can be represented categorically as the carrier of the
initial algebra for (i.e., the least fixed-point of) its signature functor $F$
\cite{Ma90_Data-Structures-Program-Transform}, with the definition of a
conventional \(F\)-algebra in type theory as a pair $(X,\phi)$, where $X$ is
a type (called the \emph{carrier}) and $\phi$ is a function of type $F ·X ➔
X$. A Mendler-style \(F\)-algebra, which can also be used to define
$D$~\cite{UV99_Mendler-Inductive-Types,
  Me91_Inductive-Types-20-Lambda-Calculus}, is a pair $(X,\Phi)$, where $X$ is
still a type but now function $\Phi$ has type $\abs{∀}{R}{★}{(R ➔ X) ➔ F ·R ➔
  X}$, with the $R ➔ X$ argument used to make recursive calls on subdata of the
quantified type $R$. A Mender-style \textit{CoV algebra} additionally equips
$\Phi$ with an abstract destructor (i.e., fixpoint unrolling function) via an
argument of type \(R ➔ F ·R\), allowing for further
case analysis on subdata at the quantified type $R$.

\subsection{\textbf{Generic library}}
Briefly, we describe the definitions of the generic library of
\cite{FDJS18_CoV-Ind} utilized for datatype elaboration, given in Figures
\ref{sfig:elab-toolkit} and \ref{sfig:elab-mendler}.
To improve readability, we informally present some of these definitions as type
inference rules rather than verbatim Cedille code.
All such definitions
are definable in Cedille 1.0.0, which lacks datatypes.

\paragraph{Type coercions}
For any types $A$ and $B$, \(\texttt{Cast} ·A ·B\) is the type of generalized
identity functions in CDLE, first described by \cite{FBS18_Efficient-Mendler}.
Since CDLE is Curry-style, such a function might exist even if 
$A$ and $B$ are inconvertible. It is introduced with \(\texttt{intrCast}\
f\ p\) assuming $\ann{f}{A ➔ B}$ and $p$ is a proof that $f$ behaves
\emph{extensionally} like the identity function; if \(\ann{c}{\texttt{Cast} ·A ·B}\),
then $c$ can be eliminated with \(\texttt{elimCast}\ \mhyph c\) which has type
\(A ➔ B\) and which, crucially, is \emph{intensionally} equal (indicated by
notation $\cong$ in the figure) to $\absu{λ}{x}{x}$. 

\paragraph{Positive type schemes}
Given $\ann{F}{★ ➔ ★}$, \(\texttt{Mono} ·F\) is the type of proofs that $F$ is
positive (or \emph{monotonic}). In fact, datatype elaboration produces just such
a proof when checking positivity of a datatype declaration (Section
\ref{ssec:positivity-checker}). It is introduced by \texttt{intrMono}, which
takes as argument some $f$ of similar type to the usual lifting of a function over
a functor, but restricted to \texttt{Cast}s; if
\(\ann{im}{\texttt{Mono} ·F}\) and $\ann{c}{\texttt{Cast} ·A ·B}$, then
$\mathit{im}$ is eliminated with \(\texttt{elimMono}\ \mhyph \mathit{im}\ 
\mhyph c\) which has type \(F ·A ➔ F ·B\) and which is equal to
$\absu{λ}{x}{x}$.

\paragraph{Type fixpoints}
The type \(\texttt{Fix} ·F\ \mathit{im}\) is the least fixpoint of a type scheme
$F$ whose positivity is proven by $\ann{im}{\texttt{Mono} ·F}$. Functions
\texttt{in} and \texttt{out} are the expected \emph{rolling} and
\emph{unrolling} functions representing resp. a generic collection of 
constructors and destructors for a datatype with signature $F$.

\paragraph{Induction}
In Figure \ref{sfig:elab-mendler} we give the type signature for the induction
principle of the generic library (notice module parameters $F$ and
$\mathit{im}$).
The type
\texttt{D} (the datatype whose signature is $F$) is simply an abbreviation for
\(\texttt{Fix}\ ·F\ \mathit{im}\). The type family \texttt{PrfAlg} is a
dependent version of the Mendler-style CoV algebra. Its additional (erased)
arguments are $c$, an type coercion from $R$ to \texttt{D}, and
$\mathit{oeq}$, a proof that the abstract destructor $o$ is equal to
\texttt{out}. Argument $c$ is required to be even able to state the codomain of
\texttt{PrfAlg}, which is the type of proofs that $P$ holds of the \texttt{in}
of $\mathit{xs}$, after coercing this using \texttt{elimMono} and $c$ to type $F
·\texttt{D}$. Finally, \texttt{induction} is the generic
  induction principle for \texttt{D}.

We conclude by stating the requirements
that are needed to show \textit{value-preservation} (Theorem
\ref{thm:value-preservation}) and the \textit{termination guarantee} (Theorem
\ref{thm:cbn-normalization}) that any λ-encoding implementing this interface
must satisfy.

\begin{requirement}
  \label{req:lambek1}
  Definitions \texttt{in} and \texttt{out} are mutual inverses. Furthermore,
  there is a constant bound such that
  for all \(\ann{xs}{\texttt{F} ·\texttt{D}}\), expression \(\texttt{out}\
  (\texttt{in}\ \mathit{xs})\) β-reduces to \(\mathit{xs}\) in a number of steps
  within that bound.
\end{requirement}

\begin{requirement}
  \label{req:inductionComp}
  For every untyped λ-expression $\mathit{a}$, there exists some term $t$ such
  that \(\texttt{induction}\ a \reduce^* t\) and that\\ \(\texttt{induction}\ a\
  (\texttt{in}\ xs) \reduce^* a\ \texttt{out}\ t\ \mathit{xs}\), for all terms $\mathit{xs}$.
\end{requirement}

\begin{requirement}
  \label{req:nontrivial-fix}
  For closed definitions of $\ann{F}{★ ➔ ★}$ and
  $\ann{\mathit{im}}{\texttt{Mono} ·F}$, there exists some closed term $t'$ of
  type \(\texttt{Fix} ·F\ \mathit{im} ➔
  \abs{Π}{x}{T_1}{T_2}\) (for some $T_1$ and $T_2$) that erases to $\absu{λ}{x}{x}$.
\end{requirement}

The first part of Requirement \ref{req:lambek1} is known as \emph{Lambek's
  lemma} \cite{Lam68_A-Fixpoint-Theorem-for-Complete-Categories}.
Requirement \ref{req:inductionComp} expresses the \textit{cancellation law} for the
initial Mendler-style CoV \(F\)-algebra (phrased differently: \texttt{induction}
computes as a course-of-values recursor for data).
Requirement \ref{req:nontrivial-fix} simply states that the
elaborations of datatypes \textit{must be functional}. All three requirements
are satisfied by the library provided by \cite{FDJS18_CoV-Ind}.

\subsection{Implementing CoV Pattern Matching}
\label{ssec:elab-interface}

\begin{figure*}[h]
{\small
\begin{alltt}
module DataInterface (F: ★ ➔ ★) \{im: Mono · F\}.
import GenericCoV ·F -im.

IsD: ★ ➔ ★ = <..>
isD: IsD ·D = <..>
toD: ∀ R: ★. ∀ _: IsD ·R. R ➔ D = <..>
toFD: ∀ R: ★. ∀ _: IsD ·R. F ·R ➔ F ·D = <..>

ByCases: (D ➔ ★) ➔ Π R: ★. IsD ·R ➔ ★ = λ P: D ➔ ★. λ R: ★. λ is: IsD ·R. Π xs: F ·R. P (in (toFD -is xs)).
mu': ∀ R: ★. ∀ is: IsD ·R. Π x: R. ∀ P: D ➔ ★. ByCases ·P ·R is ➔ P (toD -is x) = <..>

ByInd: (D ➔ ★) ➔ ★ = λ P: D ➔ ★. ∀ R: ★. ∀ is: IsD ·R. (Π x: R. P (toD -is x)) ➔ ByCases ·P ·R is.
mu: Π x: D. ∀ P: D ➔ ★. ByInd ·P ➔ P x = <..>
\end{alltt}
  }
  \caption{Interface for datatype elaboration}
  \label{fig:elab-interface}
\end{figure*}

There is a discrepancy between the facilities of the generic library given in
Figure \ref{fig:elab-generic} and the features of the surface language. A
recursive function over datatype $D$ with signature $F$ defined using
\texttt{induction} has available as assumptions $\ann{o}{R ➔ F ·R}$ (where $R$
is a type variable that has been quantified over) and $\ann{oeq}{\{o ≃
  \texttt{out}\}}$. However, it is undesirable to expose in the surface language
details such as the signature $F$ or the generic destructor
\texttt{out}; we expect to be able to work with
case trees and that there is shared syntax for pattern matching over data with
both the concrete and abstract type.

This discrepancy is bridged by a small translation layer sketched in Figure
\ref{fig:elab-interface}, which serves as the interface for datatype
elaboration.
Term definitions are omitted (indicated by \texttt{<..>}); we briefly summarize
them.

\paragraph{``Is'' witnesses}
For any type $R$, \(\texttt{IsD} ·R\) is the type of triples consisting of an
type coercion of type \(\texttt{Cast} ·R ·\texttt{D}\), a generic destructor
$\ann{o}{R ➔ \texttt{F} ·R}$, and proof \(\{o ≃ \texttt{out}\}\).
It
simply packages together some of the assumptions available in any proof by
\texttt{induction}. Term \texttt{isD} is the trivial witness of \(\texttt{isD}
·\texttt{D}\). Function \texttt{toD} (which is definitionally equal to
$\absu{λ}{x}{x}$) takes evidence of \(\texttt{IsD} ·R\)
(for some $R$) and uses this to produce a type coercion from $R$ to \texttt{D}.
These definitions correspond to \genid{Is/}{D}, \genid{is/}{D},
and \genid{to/}{D} (for a given datatype $D$) in the surface language.
Function \texttt{toFD} is not exported 
to the surface language and uses
\texttt{elimMono} to cast \(\texttt{F} ·R\) to \(\texttt{F} ·\texttt{D}\), given
some term of type \(\texttt{IsD} ·R\).

\paragraph{Proofs by cases}
Type \(\texttt{ByCases} ·P ·R\ \mathit{is}\,\) is
the type of proofs that $P$ holds by case analysis, where $\ann{P}{\texttt{D} ➔
  ★}$.
Thus, the type of
\texttt{mu'} says that for any term $x$ of type $R$ where \(\texttt{IsD} ·R\)
holds, to show $P$ holds of $x$ (after casting $x$ to \texttt{D}), it suffices
to give a proof by case analysis on $R$.
Its definition uses the abstract
destructor $o$ (given by \(\ann{is}{\texttt{IsD} ·R}\)) on $x$.
Definition \texttt{mu'} corresponds directly to \mup\, in the surface language.

\paragraph{Proofs by induction}
\(\texttt{ByInd} · P\) is the generic type of proofs that property $P$
holds by induction.
It is defined using \texttt{ByCases}
and additionally equipped with an inductive hypothesis and evidence of
\(\texttt{IsD} ·R\) for the quantified type $R$.
Thus, the type of \texttt{mu}
says that $P$ holds for any $\ann{x}{D}$ when, under the assumption that $P$
holds for every $x'$ of type $R$ (for some $R$ for which \(\texttt{IsD} ·R\) holds), $P$ holds for
the \texttt{in} of $\mathit{xs}$ for all $\ann{xs}{F ·R}$; use of the abstract type $R$
breaks circularity.
Definition \texttt{mu} corresponds to μ in the surface and uses
\texttt{induction}, repackaging the assumptions available to the
\texttt{PrfAlg} argument for use by its argument of type \(\texttt{ByInd}
·P\).

\section{Elaboration of Datatype Declarations}
\label{sec:elab-data}
\paragraph{\textbf{Notation}}
In this section we give a formal description of the elaboration of non-indexed
datatypes in Cedille to λ-encodings in Cedille 1.0.0 (which lacks datatypes);
this is also the scope of the accompanying proof appendix. A declaration
of datatype $D$ of kind $★$ is written $\indsche{D}{R}{\Delta}$, where
\begin{itemize}
\item $R$ is a fresh type-variable of kind \texttt{★} whose scope is $\Delta$
\item $\Delta$ is an association of the constructors of $D$ with their type signatures such
  that all occurrences of $D$ in the types of constructor arguments in the surface
  language have been replaced by $R$
\end{itemize}

\noindent For example, the declaration of \texttt{Nat} in Figure
\ref{fig:cedille-data-standard} translates to
\[\indsche{\texttt{Nat}}{R}{\small
    \begin{array}{lcl}
      \texttt{zero} & : & \texttt{Nat}
      \\ \texttt{suc} & : & \texttt{$R$ ➔ Nat}
    \end{array}
  }\]

We write \(\Gamma \vdash \indsche{D}{R}{\Delta}\ \wf\) to indicate that, for
every $i$ ranging from $1$ to the number of constructors in $\Delta$ (written
$i=1..\#\Delta$), the $i$th constructor $c_i$ in $\Delta$ has a type of the form
$\piforall \vars{a_i: A_i}. D$ (indicating the mixed-erasure quantification over
the dependent telescope of terms and types $\vars{\ann{a_i}{A_i}}$)
that is well-kinded in context $\Gamma$ extended by variables
$\ann{R}{★}$ and $\ann{D}{★}$, and furthermore there are no occurrences of $D$
in the classifiers of the telescope $\vars{\ann{a_i}{A_i}}$.
Notations $\lamLam\ \vars{a_i}. t$ and $t\ \vars{a_i}$ indicates resp. the
term-level abstraction and application over this telescope that respects the
erasures and classifiers over which the variables were quantified. This
convention generalizes to the sequence of term and type expressions
$\vars{s_i}$, as in \(P\ (c_i\ \vars{s_i})\), when indicated that $\vars{s_i}$ are
produced from type and kind coercions of $\vars{a_i}$.

By convention, judgments with the hooked arrow $\Gamma \vdash t : T \elab
t'$ are elaboration rules, written $\Gamma \vdash t : T \elab \_$ when we need
only that $t$ is well-typed. Judgments without hooked arrows $\Gamma \vdash t :
T$ and $\Gamma \vdash T : K$ indicate typing and kinding in Cedille 1.0.0. Some
inference rules have premises of the form \((\Gamma \vdash
\absu{\piforall}{\vars{\ann{a_i}{A_i}}}{T} : ★ \elab \_)_{i=1..\#\Delta}\),
accompanied by a premise \((c_i : \absu{\piforall}{\vars{\ann{a_i}{A_i}}}{D} \in
\Delta)_{i=1..\#\Delta}\); the first indicates a family of derivations of the
parenthesized judgment indexed by the $i$th constructor of $\Delta$ and its
constructor argument telescope \(\vars{\ann{a_i}{A_i}}\), and the second merely
names these telescopes explicitly and exhaustively. $\Gamma\splab{G}$ indicates
a typing context consisting of the definitions in Figures \ref{fig:elab-generic}
and \ref{fig:elab-interface}.

$\mathit{Italics}$ indicates meta-variables,
\texttt{teletype} font indicates code literals (except in meta-variables
denoting generated names like \genid{Is/}{D}), and $^{\text{superscript}}$
denotes labels for meta-variables.
We use the following labeling convention for expressions
elaborated from datatypes and their constructors: \splab{F} for the usual
impredicative encoding of a datatype's signature type scheme; \splab{FI} for the datatype
signature formed by dependent intersection and supporting a proof principle; and
\splab{FIX} for the least fixpoint of the datatype's ``proof signature.''

\subsection{Datatype and Constructor Elaboration}
\label{ssec:elab-datatypes}
\begin{figure*}[h!]
  \centering
  \begin{subfigure}{1\linewidth}
  \[
    \begin{array}{c}
      \infer[\textsc{[F]}]{
      \Gamma \vdash \indsche{D}{R}{\Delta}
      \elalab{F}
      \absu{λ}{R}{
       \abs{∀}{X}{★}{
        \abs{(Π}{x_i}{\absu{\piforall}{\vars{\ann{a_i}{A_i'}}}{X})_{i=1..\#\Delta}}{X}
       }
      }
      }{
      (\ann{c_i}{\absu{\piforall}{\vars{\ann{a_i}{A_i}}}{D}} \in \Delta)_{i=1..\#\Delta}
      \quad (\Gamma,\ann{R}{★},\ann{X}{★}
      \vdash \absu{\piforall}{\vars{\ann{a_i}{A_i}}}{X} : ★
      \elab \absu{\piforall}{\vars{\ann{a_i}{A_i'}}}{X})_{i=1..\#\Delta}
      }

      \\ \\
      \infer[\textsc{[cF]}]{
      \Gamma \vdash (\indsche{D}{R}{\Delta}, j)
      \elalab{cF}
      \absu{Λ}{R}{
       \absu{\lamLam}{\vars{a_j}}{
        \absu{Λ}{X}{
         \absu{λ}{x_{i=1..\#\Delta}}{x_j\ \vars{a_j}}
        }
      }
      }
      }{
      \ann{c_j}{\absu{\piforall}{\vars{\ann{a_j}{A_j}}}{D}} \in \Delta
        }

      \\ \\
      \infer[\textsc{[FI]}]{
      \begin{array}{c}
        \Gamma \vdash \indsche{D}{R}{\Delta}
        \elalab{FI}
        \absu{λ}{R}{
         \abs{ι}{x}{D\splab{F} ·R}{
          \abs{∀}{X}{D\splab{F} ·R ➔ ★}{
            \abs{(Π}{x_i}{\absu{\piforall}{\vars{\ann{a_i}{A_i'}}}{X\
             (c\splab{F}_i\ \vars{a_i})})_{i=1..\#\Delta}}{X\ x}
          }
         }
        }
      \end{array}
      }{
      \begin{array}{cc}
        \Gamma \vdash \indsche{D}{R}{\Delta} \elalab{F} D\splab{F}
        &
          (\Gamma \vdash (\indsche{D}{R}{\Delta},i)
          \elalab{cF}\ c\splab{F}_i)_{i=1..\#\Delta}
        \\ [3pt]
        (\ann{c_i}{\absu{\piforall}{\vars{\ann{a_i}{A_i}}}{D}} \in \Delta)_{i=1..\#\Delta}
        &
          (\Gamma,\ann{R}{★},\ann{X}{★}
          \vdash \absu{\piforall}{\vars{\ann{a_i}{A_i}}}{X} : ★
          \elab \absu{\piforall}{\vars{\ann{a_i}{A_i'}}}{X})_{i=1..\#\Delta}
      \end{array}
      }

      \\ \\
      \begin{array}{cc}
        \infer[\textsc{[cFI]}]{
        \Gamma \vdash (\indsche{D}{R}{\Delta},j)
        \elalab{cFI}
        \absu{Λ}{R}{\absu{\lamLam}{\vars{a_j}}{
        [c\splab{F}_j\ \vars{a_j}
        , \absu{Λ}{X}{\absu{λ}{x_{i=1..\#\Delta}}{x_j\ \vars{a_j}}}]
        }}
        }{
        \Gamma \vdash (\indsche{D}{R}{\Delta},j) \elalab{cF} c^F_j
        \quad \ann{c_j}{\absu{\piforall}{\vars{\ann{a_j}{A_j}}}{D}} \in \Delta
        }
        &
          \infer[\textsc{[FIX]}]{
          \Gamma \vdash \indsche{D}{R}{\Delta} \elalab{FIX}\
          \texttt{Fix} ·D\splab{FI}\ pos
          }{
          \Gamma \vdash \indsche{D}{R}{\Delta} \elalab{FI}\ D\splab{FI}
          \quad \Gamma \vdash D\splab{FI} \elapos pos
          }
      \end{array}

      \\ \\
      \infer[\textsc{[cFIX]}]{
      \Gamma \vdash (\indsche{D}{R}{\Delta}, j)
      \elalab{cFIX}
      \absu{\lamLam}{\vars{a_j}}{
       \ \texttt{in} ·D\splab{FI}\ \mhyph pos\ (c\splab{FI}_j·(\texttt{Fix}
      ·D\splab{FI}\ pos)\ \vars{a_j})
      }
      }{
      \Gamma \vdash \indsche{D}{R}{\Delta} \overset{\text{FIX}}{\elab}
      \texttt{Fix} ·D\splab{FI}\ pos
      \quad \Gamma \vdash (\indsche{D}{R}{\Delta},j)
      \overset{\text{cFI}}{\elab}\ c\splab{FI}_j
      \quad \ann{c_j}{\absu{\piforall}{\vars{\ann{a_j}{A_j}}}{D}} \in \Delta
      }

      \\ \\
      \infer[\textsc{[Data]}]{
      \Gamma \vdash \indsche{D}{R}{\Delta}
      \dashv \Gamma, \indel{D}{R}{\Delta}{\Theta}{\mathcal{E}}
      }{
      \begin{array}{c}
        \Gamma \vdash \indsche{D}{R}{\Delta}\ \mathit{wf}
        \quad \Gamma \vdash \indsche{D}{R}{\Delta} \elalab{FIX} \texttt{Fix} ·D\splab{FI}\ pos
        \quad (\Gamma \vdash (\indsche{D}{R}{\Delta}, i)
        \elalab{cFIX} c\splab{FIX}_i)_{i=1..\#\Delta}
        \\[2pt]
        \Theta = (\texttt{Is/$D$} : ★ ➔ ★,\
         \texttt{is/$D$} : \texttt{Is/}D ·D,\
         \texttt{to/$D$} = \absu{λ}{x}{x}
          : \abs{∀}{R}{★}{\abs{∀}{is}{\texttt{Is/}D ·R}{R ➔ D}})
        \\[2pt]
        \mathcal{E} = \left\{
        \begin{array}{c}
          D \mapsto \texttt{Fix} ·D\splab{FI}\ pos,
          \quad (c_i \mapsto c_i\splab{FIX})_{i=1..\#\Delta},
          \\[2pt] \texttt{Is/$D$} \mapsto \texttt{IsD} ·D\splab{FI}\ pos,
          \quad \texttt{is/$D$} \mapsto \texttt{isD} ·D\splab{FI}\ \mhyph pos,
          \quad \texttt{to/$D$} \mapsto \texttt{toD} ·D\splab{FI}\ \mhyph pos
        \end{array}
        \right\}
      \end{array}
      }
    \end{array}
  \]
  \end{subfigure}
  \caption{Elaboration of datatype declarations}
  \label{fig:elab-data-decl}
\end{figure*}

Figure \ref{fig:elab-data-decl} shows elaboration of a datatype $D$ and its
constructors. To improve readability we give a \textit{set of judgments}, each
formed from a \textit{single rule} performing one task: \textsc{[F]} and \textsc{[cF]} elaborate
resp. the usual impredicative encoding of a datatype's signature type scheme and its
constructors; \textsc{[FI]} and \textsc{[cFI]} the inductive signature and its
constructors; \textsc{[FIX]} and \textsc{[cFIX]} the least fixpoint of the
inductive signature and its constructors; and \textsc{[Data]} of the form \(\Gamma \vdash
\indsche{D}{R}{\Delta} \dashv \Gamma,\indel{D}{R}{\Delta}{\Theta}{\mathcal{E}}\)
adds the datatype, constructors, globals ($\Theta$), and elaborations
($\mathcal{E}$) to the context.

In rule \textsc{[F]}, the first premise serves to name the family of constructor
argument telescopes \((\vars{\ann{a_i}{A_i}})_{i=1..\#\Delta}\), and the second
premise elaborates the family of types
\(\absu{\piforall}{\vars{\ann{a_i}{A_i}}}{X}\), where $X$ is fresh wrt $\Gamma$.
The body of the elaborated type scheme is a function type quantifying over $X$
and abstract constructors $x_i$ for $i=1..\#\Delta$ (themselves functions
quantifying over the appropriate elaborated constructor argument types) with
codomain $X$. In rule \textsc{[cF]} we elaborate the $j$th constructor for this
signature type scheme, abstracting over the recursive-occurrence type $R$, the $j$th
sequence of arguments \(\vars{a_j}\), and abstract constructors $x_i$ to produce
$x_j\ \vars{a_j}$ Concretely, the elaborations for \texttt{Nat} by these two rules
are:%
{ \small
\begin{alltt}
Nat\(\splab{F}\): ★ ➔ ★ = λ R: ★. ∀ X: ★. Π z: X. Π s: R ➔ X. X.
zero\(\splab{F}\): ∀ R: ★. Nat\(\splab{F}\) ·R = Λ R. Λ X. λ z. λ s. z.
suc\(\splab{F}\): ∀ R: ★. R ➔ Nat\(\splab{F}\) ·R = Λ R. λ n. Λ X. λ z. λ s. s n.
\end{alltt}
}

The next two rules, \textsc{[FI]} and \textsc{[cFI]}, show elaboration to resp.
the inductive signature type scheme and its constructors.
The type scheme
elaborated by \textsc{[FI]} returns from a type argument $R$ the \textit{dependent
intersection} of $\ann{x}{D\splab{F} ·R}$ (where $D\splab{F}$ is produced by rule
\textsc{[F]}) and a proof that, for any property \(\ann{X}{D\splab{F} ·R ➔ ★}\), $X\
x$ holds if $X$ holds for the constructors of
$D\splab{F} ·R$ applied to their arguments (\(X\ (c\splab{F}_i\ \vars{a_i})\) in
the rule). \textsc{[cFI]} elaborates the $j$th constructor of the inductive
signature $D\splab{FI}$, whose first component 
$c\splab{F}_j\ \vars{a_j}$ is the $j$th constructor of \(D\splab{F} ·R\) applied to its
arguments and whose second component is a proof (by using the appropriate assumption
$x_j$) that \(X\ (c\splab{F}_j\ \vars{a_j})\) holds. The two components
are indeed convertible (modulo erasure), satisfying the requirements for
introducing a dependent intersection.
Concretely, the elaborations for \texttt{Nat} by these rules are:
{\small
\begin{alltt}
Nat\(\splab{FI}\): ★ ➔ ★ = λ R: ★.
  ι x: Nat\(\splab{F}\) ·R. ∀ X: Nat\(\splab{F}\) ·R ➔ ★.
       Π z: X zero\(\splab{F}\). Π s: (Π r: R. X (suc\(\splab{F}\) r)). X x.

zero\(\splab{FI}\): ∀ R: ★. Nat\(\splab{FI}\) ·R
= Λ R. [zero\(\splab{F}\) ·R, Λ X. λ z. λ s. z].

suc\(\splab{FI}\): ∀ R: ★. R ➔ Nat\(\splab{FI}\) ·R
= Λ R. λ n. [suc\(\splab{F}\) n , Λ X. λ z. λ s. s n].
\end{alltt}
} Rules \textsc{[FIX]} and \textsc{[cFIX]} tie the recursive knot using the
generic interface of Figure \ref{fig:elab-interface}: datatype $D$ elaborates to
\(\texttt{Fix} ·D\splab{FI}\ pos\), where $D\splab{FI}$ is produced by {\small
  \textsc{[FI]}} and $pos$ is a term of type \texttt{Mono ·$D\splab{FI}$}
(i.e., a proof that $D\splab{FI}$ is covariant) whose production is described in
Section \ref{ssec:positivity-checker}. Rule \textsc{[cFIX]} elaborate datatype constructors to
the \texttt{in} of the constructors of $D\splab{FI}$ applied to their arguments
and instantiated to type \(\texttt{Fix} ·D\splab{FI}\ pos\). Finally, rule
\textsc{[Data]} associates the datatype declaration with its
elaboration in the typing context, with $\Theta$ binding the globals
\genid{Is/}{D}, \genid{is/}{D}, and \genid{to/}{D} and
$\mathcal{E}$ associating datatype $D$, its constructors, and its globals with
their elaborations. Note that \genid{to/}{D} in $\Theta$ is \emph{defined}
to be $\absu{λ}{x}{x}$ (not just declared to have a type) for purposes
of definitional equality.

\paragraph{\textbf{Soundness Properties}}
The elaborations of datatype declarations enjoys the following soundness property.

\begin{theorem}[Elaboration of declarations]
  \label{thm:elab-data-sound}

  Assuming
  \begin{itemize}
  \item \(\Gamma \vdash \indsche{D}{R}{\Delta} \dashv
    \Gamma,\indel{D}{R}{\Delta}{\Theta}{\mathcal{E}}\), and \\
    \((\ann{c_i}{\absu{\piforall}{\ann{\vars{a_i}}{A_i}}{D}} \in
    \Delta)_{i=1..\#\Delta}\)

    (we have elaborated a well-formed datatype with constructors of a
    certain shape)
  \item \(\vdash \Gamma \elab \Gamma'\) (the typing context
    elaborates, Figure \ref{fig:elab-all})
  \item \((\Gamma,\ann{X\!}{★},\ann{R\!}{★} \vdash
    \absu{\piforall}{\vars{\ann{a_i}{A_i}}}{X} : ★ \elab
    \absu{\piforall}{\vars{\ann{a_i}{A_i{'}}}}{X}\)\, implies
    \(\Gamma\splab{G},\Gamma',\ann{R}{★},\ann{X}{★} \vdash
    \absu{\piforall}{\vars{\ann{a_i}{A_i{'}}}}{X} : ★\))$_{i=1..\#\Delta}$

    (the elaborated constructor argument types are well-kinded)
  \end{itemize}
  we have that
  \begin{itemize}
  \item \(\Gamma\splab{G},\Gamma' \vdash \mathcal{E}(D) : ★\) and \\
    \((\Gamma\splab{G},\Gamma' \vdash \mathcal{E}(c_i) :
    \absu{\piforall}{\vars{\ann{a_i}{[\mathcal{E}(D)/R]A_i'}}}{\mathcal{E}(D)})_{i=1..\#\Delta}\)

    (the elaborated datatype and constructors have the expected kind and
    type)
  \item \(\Gamma\splab{G},\Gamma' \vdash \mathcal{E}(\genid{Is/}{D}) : ★
    ➔ ★\),

    \(\Gamma\splab{G},\Gamma' \vdash \mathcal{E}(\genid{is/}{D}) :
    \mathcal{E}(\genid{Is/}{D}) · \mathcal{E}(D)\)

    (the elaborations of \genid{Is/}{D} and \genid{is/}{D} have their
    expected kinds and types)
  \item \(\Gamma\splab{G},\Gamma' \vdash \mathcal{E}(\genid{to/}{D}) :
    \abs{∀}{R}{★}{
      \abs{∀}{is}{\mathcal{E}(\genid{Is/}{D}) ·R}{R ➔ \mathcal{E}(D)}
    }\), with \(|\mathcal{E}(\genid{to/}{D})| =_{\beta\eta} \absu{λ}{x}{x}\)

    (the elaboration of $\genid{to/}{D}$ has its expected type and convertibility)
  \end{itemize}
\end{theorem}

\subsection{Positivity Checker}
\label{ssec:positivity-checker}
\begin{figure*}[h!]
  \begin{subfigure}{1\linewidth}
    \caption{Main judgments and positivity rule}
  \[
    \begin{array}{c}
      \begin{array}{cccc}
        \fbox{\(\Gamma \vdash F \elapos \mathit{pos}\)}
        & \fbox{\(\Gamma\, ;\, s \vdash S \les T \elales s'\)}
        & \fbox{\(\Gamma\, ;\, s \vdash K_1 \les K_2 \elales S\)}
        & \fbox{\(\Gamma\, ;\, s \vdash (\vars{\ann{a}{A_1}}) \les (\vars{\ann{a}{A_2}})
          \elab \vars{s}\)}
        \\ Positivity & Subtyping & Subkinding & Telescope\ coercion
      \end{array}
      \\ \\
      \infer[]{
       \Gamma \vdash \abs{λ}{R}{★}{T}
       \elapos
       \texttt{intrMono}\ 
        (\absu{Λ}{R_1}{\absu{Λ}{R_2}{\absu{λ}{z}{\texttt{intrCast}\ s'\ (\absu{λ}{\_}{β}})}})
      }{
      \Gamma,\ann{R}{★} \vdash T : ★ \elab \_
      \quad
      \Gamma,\ann{R_1}{★},\ann{R_2}{★},\ann{z}{\texttt{Cast} ·R_1 ·R_2}\ ;\ \texttt{elimCast}\
      \mhyph z \vdash [R_1/R]T \les [R_2/R]T \elales s'
      }
      \\ \\
    \end{array}
  \]
\end{subfigure}
\begin{subfigure}{1.0\linewidth}
  \caption{Subtyping rules (incomplete listing)}
  \label{sfig:elab-subtyping}
  \[
    \begin{array}{c}
    \begin{array}{cc}
      \infer{
       \Gamma; s \vdash S \les T \elales s
      }{
       \Gamma \vdash s : S ➔ T \elab \_
       \quad |s| \cong \absu{λ}{x}{x}
      }
      &
      \infer{
        \Gamma ; s' \vdash \abs{Π}{x}{S_1}{T_1} \les \abs{Π}{x}{S_2}{T_2}
        \elales
        \absu{λ}{f}{\absu{λ}{y}{t\ (f\ (s\ y))}}
        }{
        \Gamma; s' \vdash S_2 \les S_1 \elales s
        \quad \Gamma,\ann{y}{S_2} ; s' \vdash
        [(s\ y)/x]T_1 \les [y/x]T_2 \elales t
        }
    \end{array}
    \end{array}
  \]
\end{subfigure}
  \caption{Positivity checker}
  \label{fig:elab-positivity}
\end{figure*} 

Figure \ref{fig:elab-positivity} lists the judgments used for checking datatype
positivity, the single rule defining its primary judgment \(\Gamma \vdash
\abs{λ}{R}{★}{T} \elapos \mathit{pos}\) that proves $\abs{λ}{R}{★}{T}$ (of kind
$★ ➔ ★$) is positive, and some representative rules for the subtyping judgment;
the complete set of rules is given in the proof appendix.
These rules are
\textit{evidence-producing}, as the elaborator interface of Section
\ref{sec:generic-interface} (specifically rule \textsc{[FIX]}) requires explicit proof of positivity in the form of
\texttt{Mono}.
This proof is generated by invoking the
subtyping judgment, with an intuitive reading that
$\abs{λ}{R}{★}{T}$ is positive if for any $R_1$ and $R_2$ where $R_1 \les R_2$
we have $[R_1/R]T \les [R_2/R]T$ (with $\les$ suggesting a form of subtyping whose
semantics is \texttt{Cast}).

In the subtyping judgment \(\Gamma ; s \vdash S \les T \elales s'\), 
input $s$ witnesses a base subtyping assumption (demonstrated in Figure
\ref{sfig:elab-subtyping}, top left), and output $s'$ is a coercion derived from it,
definitionally equal to $\absu{λ}{x}{x}$.
In the positivity rule, this input is \(\texttt{elimCast}\ \mhyph z\) for an arbitrary
$z$ of type $\texttt{Cast} ·R_1 ·R_2$, and the output has type $[R_1/R]T ➔ [R_2/R]T$.
To illustrate the machinery of the subtyping judgment, consider the rule for
Π-types (top right): to show \(\abs{Π}{x}{S_1}{T_1} \les \abs{Π}{x}{S_2}{T_2}\)
with base assumption $s'$, first produce for the domain a
coercion $s$ proving $S_2 \les S_1$ (note the contravariance), then produce for
the codomain a
coercion $t$ proving for all $\ann{y}{S_2}$, \([(s\ y)/x]T_1 \les [y/x]T_2\);
the coercion in the conclusion clearly \(\beta\eta\)-reduces to $\absu{λ}{f}{f}$
since coercions $s$ and $t$ do.

A similar reading as for subtyping holds for the subkinding judgment
$\Gamma ; s \vdash K_1 \les K_2 \elales S$, though the shape of kind coercion
$S$ need not be specified (all types are erased in terms).
The telescope
coercion judgment breaks the pattern by producing a coerced sequence of terms
and types $\vars{s}$ (and not the coercions) from a telescope
\(\vars{\ann{a}{A_1}}\).
These are equal (modulo erasure) to \(\vars{a}\) and
typeable with telescope \(\vars{\ann{a}{A_2}}\) (Figure \ref{fig:elab-all});
they are used in Figure \ref{fig:elab-mu} to state the expected type of each
case body for μ- and {\mup}-expressions.

This description of our positivity checker is made precise
by the following soundness properties:
\begin{theorem}[Positivity checker]
  \label{thm:elab-positivity}
  \
  \begin{enumerate}
  \item If\, \(\Gamma ; s \vdash S \les T \elales s'\) then\, $\Gamma \vdash s' :
    S ➔ T \elab \_$ and $|s'| =_{\beta\eta} \absu{λ}{x}{x}$
  \item If\, \(\Gamma ; s \vdash K_1 \les K_2 \elales S\) then\, \(\Gamma \vdash
    S : K_1 ➔ K_2 \elab \_\)
  \item If\, \(\Gamma \vdash F \elapos \mathit{pos}\) then\, $\Gamma,\Gamma\splab{G} \vdash \mathit{pos} :
    \texttt{Mono} ·F \elab \_$
    
  \item
    If\, \(\Gamma ; s' \vdash \vars{\ann{a}{A}} \les \vars{\ann{a}{B}}
    \elales \vars{s}\)
    then \(\Gamma\, ;\, (\vars{\ann{a}{A}}) \vdash \vars{s} : (\vars{\ann{a}{B}})
    \elab \_\)
    and \(|\vars{s}| \cong |\vars{a}|\)
  \end{enumerate}
\end{theorem}

These properties are self-explanatory, except for (4) which makes use of a new
judgment form \(\Gamma\,;\, (\vars{\ann{a}{A}}) \vdash \vars{s} :
(\vars{\ann{a}{B}}) \elab \_\) (Figure \ref{fig:elab-all}). This judgment is
read ``under $\Gamma$ and a telescope $\vars{\ann{a}{A}}$, the sequence
$\vars{s}$ is classified by the telescope $\vars{\ann{a}{B}}$'', and is defined
by progressively extending the context by each variable in $\vars{\ann{a}{A}}$,
typing (kinding) each term (type) in $\vars{s}$ according to each classifier
in \(\vars{\ann{a}{B}}\), and substituting this term (type) into
the remainder of the telescope \(\vars{\ann{a}{B}}\).

\section{Elaboration of Datatype Functions}
\label{sec:datatype-functions}
This section details the typing, operational semantics, and elaboration of μ-
and {\mup}-expressions.
Due to space restrictions we save for the separate proof appendix the complete
listing of elaboration rules, as elaborating the rest of Cedille is
straightforward: all occurrences of datatypes, their constructors, and exported
global definitions are replaced with the elaborations mapped by $\mathcal{E}$
(Figure ~\ref{fig:elab-data-decl}), auxiliary rules for elaborating the context
and type-coerced constructor arguments, and congruence rules for elaborating
non-datatype term, type, and kind constructs.
The main judgments comprising elaboration are given in Figure~\ref{fig:elab-all}.
The elaboration rules are made syntax-directed with a \emph{bidirectional} reading
\cite{PT00_Local-Type-Inference} wherein types for elimination forms (such as μ
and {\mup}) are \emph{checked} and those for introduction forms (such as
constructors, not shown) are \emph{synthesized}.
\subsection{Type inference rules}
\label{ssec:elab-mu}
\begin{figure}[ht!]
  {\small
\begin{alltt}
import DataInterface ·\(D\splab{FI}\)\ -\(pos\).
Lift\(\sb{D}\) : Π P: \(D\splab{FIX}\) ➔ ★. Π R: ★. Π is: IsD ·R. \(D\splab{F}\) ·R ➔ ★
= λ P: \(D\splab{FIX}\) ➔ ★. λ R: ★. λ is: IsD ·R. λ x: \(D\splab{F}\) ·R.
  ∀ m: \(D\splab{FI}\) ·R. ∀ eq: \{m ≃ x\}.
  P (in (φ eq - (toFD -is m) \{x\})).
\end{alltt}
  }
  \caption{Lifting of properties of $D\splab{FIX}$ to $D\splab{F}$}
  \label{fig:elab-lift}
\end{figure}

\paragraph{\textbf{Property lifting}}
The elaboration rules in Figure \ref{fig:elab-data-decl} are able to satisfy the
elaborator interface of Figure \ref{fig:elab-interface} by producing from a
well-formed declaration of positive datatype $D$ a signature functor
$D\splab{FI}$ and positivity proof $\mathit{pos}$. Even so, it is not yet
obvious that the appropriate arguments to functions \texttt{mu'} and \texttt{mu}
can be given when elaborating \mup- and μ-expressions. In particular, both
\texttt{mu'} and \texttt{mu} require proofs \texttt{ByCases}, which in general
requires a (non-recursive) dependent eliminator for
$D\splab{FI} ·D\splab{FIX}$. The careful reader will have noted in the preceding
section that type scheme $D\splab{FI}$ produced by rule \textsc{[FI]}
\textit{does} support proofs by cases -- but only for properties stated over the
type scheme $D\splab{F}$ produced by {\small [F]}.

The solution to this mismatch is given in Figure \ref{fig:elab-lift}, listing
the type-level function \texttt{Lift$_D$} \textit{lifting properties $P$ of
  $D\splab{FIX}$ to a property of $D\splab{F}$}. Given such $P$, a type $R$, a
witness $\mathit{is}$ of type \(\texttt{IsD} ·R\), and $x$ of type $D\splab{F} ·R$,
\(\texttt{Lift}_D ·P ·R\ \mathit{is}\ x\) is a proof that $P$ holds for the \texttt{in}
of $x$, where the φ-expression casts $x$ to the type $D\splab{FI} ·D\splab{FIX}$
of the expression \(\texttt{toFD}\ \mhyph \mathit{pos}\ \mhyph \mathit{is}\ m\), for any $m$ equal
(by $\mathit{eq}$) to $x$. Recall that \texttt{toFD} erases to $\absu{λ}{x}{x}$;
thus the expression \(\texttt{toFD}\ \mhyph \mathit{pos}\ \mhyph \mathit{is}\
m\) is convertible with $m$. Because of this, $\mathit{eq}$
really does prove these two expressions are equal.

\begin{figure*}
  \begin{subfigure}{1.0\linewidth}
    \[
      \begin{array}{cccccccc}
        \fbox{\(\Gamma \vdash t : T \elab t'\)}
        & \fbox{\(\Gamma \vdash T : K \elab T'\)}
        & \fbox{\(\Gamma \vdash K \elab K'\)}
        & \fbox{\(\Gamma \vdash t \elab p\)}
        & \fbox{\(\vdash \Gamma \elab \Gamma{'}\)}
        & \fbox{\(\Gamma\,;\, (\vars{\ann{a}{A}}) \vdash \vars{s} :
          (\vars{\ann{a}{B}}) \elab \vars{s'}\)}
        \\ \mathit{Terms} & \mathit{Types} & \mathit{Kinds}
        & \mathit{Pure\ terms} & \mathit{Contexts} & \mathit{Telescope\ coercions}
      \end{array}
    \]
  \end{subfigure}
  \caption{Elaboration judgments}
  \label{fig:elab-all}
\end{figure*}

\begin{figure*}
  \begin{subfigure}{1.0\linewidth}
    \[\infer[\textsc{[Cases]}]{ \Gamma \vdash \{c_i\ \vars{a_i} ➔
        t_i\}_{i=1..\#\Delta} : \text{Cases}(\{P\ (c_i\
        \vars{s_i})\}_{i=1..\#\Delta},\mathit{is}) \elab
        (\{\absu{\lamLam}{\vars{a_i}}{t_i{'}}\}_{i=1..\#\Delta}, \mathit{is'})
      }{
        \begin{array}{c}
          \Gamma \vdash \mathit{is} : \genid{Is/}{D} ·T \elab \mathit{is'}
          \quad \indel{D}{R}{\Delta}{\Theta}{\mathcal{E}} \in \Gamma
          \quad \genid{Is/}{D} \in \Theta,\genid{to/}{D} \in \Theta
          \quad (c_i : \piforall \vars{a_i: A_i}. D \in \Delta)_{i=1..\#\Delta}
          \\ [2pt]
          (\Gamma; \genid{to/}{D}\ \mhyph \mathit{is}
          \vdash \vars{\ann{a_i}{[T/R]A_i}} \les \vars{\ann{a_i}{[D/R]A_i}}
          \elales \vars{s_i})_{i=1..\#\Delta}
          \quad (\Gamma \vdash \absu{\lamLam}{\vars{a_i}}{t_i}
          : \absu{\piforall}{\vars{\ann{a_i}{[T/R]A_i}}}{P\ (c_i\
          \vars{s_i})}
          \elab \absu{\lamLam}{\vars{a_i}}{t_i'})_{i=1..\#\Delta}
        \end{array}
      }
    \]
  \end{subfigure}
  \caption{Elaboration of case branches}
  \label{fig:elab-cases}
\end{figure*}

\paragraph{\textbf{Case branches}}
To aid in reading the elaboration rules for μ and \mup-expressions, we separate
into a single judgment the book-keeping common to both for elaborating
constructor case branches. The single rule \textsc{[Cases]} forming this judgment is
given in Figure \ref{fig:elab-cases}. It should be read as taking as input a
case tree \(\{c_i\ \vars{a_i} ➔ t_i\}_{i=1..\#\Delta}\), a type family $P$, and
a witness $\mathit{is}$, and producing type-coerced constructor arguments
$\vars{s_i}$, a collection
\(\{\absu{\lamLam}{\vars{a_i}}{t_i'}\}_{i=1..\#\Delta}\) of elaborated case
bodies, and an elaborated witness $\mathit{is'}$. In its premises, we check that
the given witness $\mathit{is}$ has type $\genid{Is/}{D} ·T$ (where
\(\genid{Is/}{D}\) is associated with some declared datatype $D$) and elaborate
it, then check that constructors of the case tree cover exhaustively the
constructors of $D$ (and are given the correct number of and erasures for the
pattern-bound variables). We produce $\vars{s_i}$ via telescope coercion of $\vars{a_i}$ (Figure
\ref{fig:elab-positivity}), using the coercion \(\genid{to/}{D}\ \mhyph
\mathit{is}\) to cast recursive occurrences of $T$ (given by the occurrences
$R$) to $D$ in the types of pattern-bound variables given to $c_i$ in the case
tree. In the last premise, we elaborate each case branch at its expected
type -- a mixed-erasure abstraction over the constructor arguments $\vars{a_i}$
with codomain $P\ (c_i\ \vars{s_i})$.

\begin{figure*}
  \begin{subfigure}{1\linewidth}
    \[
      \begin{array}{c}
        \infer[\textsc{[Mu]}]{
        \begin{array}{c}
          \Gamma \vdash \mufix{ih}{t}{P}{c_i\ \vars{a_i} ➔ t_i}_{i=1..\#\Delta}
          : P\ t
          \elab
          \texttt{mu} ·D\splab{FI}\ \mhyph pos\ t' ·P'\ 
          \\
          (\absu{Λ}{\genid{Type/}{ih}}{
           \absu{Λ}{\genid{isType/}{ih}}{
            \absu{λ}{ih}{
              \absu{λ}{x}{
                x.2 ·(\texttt{Lift}_D ·P' ·\genid{Type/}{ih}\ \genid{isType/}{ih})\ 
                 (\absu{\lamLam}{\vars{a_i}}{\absu{Λ}{m}{\absu{Λ}{eq}{t_i'}}})_{i=1..\#\Delta}\ 
                 \mhyph x\ \mhyph β
              }
            }
           }
          })
        \end{array}
        }{
        \begin{array}{c}
          \indel{D}{R}{\Delta}{\Theta}{\mathcal{E}} \in \Gamma, \mathcal{E}(D) =
          \texttt{Fix} ·D\splab{FI}\ pos,
          \genid{to/}{D} \in \Theta
          \quad \Gamma \vdash P : D ➔ ★ \elab P'
          \quad \Gamma \vdash t : D \elab t'
          \\[2pt]
          \Gamma' = \Gamma, \genid{Type/}{ih} : ★,
          \genid{isType/}{ih} : \genid{Is/}{D} ·\genid{Type/}{ih},
          \mathit{ih} : \abs{Π}{y}{\genid{Type/}{ih}}{P\ (\genid{to/}{D}\ \mhyph
          \genid{isType/}{ih}\ y)}
          \\[2pt]
          \Gamma' \vdash \{c_i\ \vars{a_i} ➔ t_i\} : \text{Cases}(\{P\ (c_i\
          \vars{s_i})\},\genid{isType/}{ih})
          \elab (\{\absu{\lamLam}{\vars{a_i}}{t_i{'}}\},\genid{isType/}{ih})
        \end{array}
        }

        \\ \\ [5pt]
        \infer[\textsc{[Mu']}]{
        \begin{array}{c}
          \mumat{is}{t}{P}{c_i\ \vars{a_i} ➔ t_i}_{i=1..\#\Delta} : P\ (\genid{to/}{D}\ \mhyph \mathit{is}\ t)
          \elab
          \texttt{mu'} ·D\splab{FI}\ \mhyph \mathit{pos}\ \mhyph \mathit{is'}\
          t' ·P'\
          \\
            (\absu{λ}{x}{\ 
             x.2 ·(\texttt{Lift}_D ·P'\ ·T{'}\ \mathit{is'})\
              (\absu{\lamLam}{\vars{a_i}}{\absu{Λ}{m}{\absu{Λ}{eq}{\ t_i'}}})_{i=1..\#\Delta}\ \mhyph x\ \mhyph β
          })
        \end{array}
        }{
        \begin{array}{c}
          \indel{D}{R}{\Delta}{\Theta}{\mathcal{E}} \in \Gamma,\
          \mathcal{E}(D) = \texttt{Fix} ·D\splab{FI}\ pos,
          \genid{to/}{D} \in \Theta 
          \quad \Gamma \vdash t : T \elab t'
          \quad \Gamma \vdash T : ★ \elab T{'}
          \quad \Gamma \vdash P : D ➔ ★ \elab P'
          \\ [2pt]
          \Gamma \vdash \{c_i\ \vars{a_i} ➔ t_i\} : \text{Cases}(\{P\ (c_i\
          \vars{s_i})\},\mathit{is}) \elab
          (\{\absu{\lamLam}{\vars{a_i}}{t_i{'}}\}, \mathit{is'})
          \end{array}
        }
      \end{array}
    \]
    \end{subfigure}
    \caption{\fbox{\(\Gamma \vdash t : T \elab t'\)} Elaboration of terms
      (shown: μ, \mup)}
    \label{fig:elab-mu}
\end{figure*}

\paragraph{\textbf{Elaboration of μ- and \mup-expressions}}
With \texttt{Lift$_D$} and rule \textsc{[Cases]} we are now able to explain how μ- and
{\mup}-expressions are elaborated, shown in Figure \ref{fig:elab-mu}.

In the premises of rule \textsc{[Mu]}, we begin by requiring that the kind of the motive
$P$ is $D ➔ ★$, and the type of the scrutinee $t$ is some concrete datatype $D$,
elaborating them to resp. $P'$ and $t'$. We then declare an extended typing
context $\Gamma'$, formed by $\Gamma$ and the μ-locals \genid{Type/}{ih},
\genid{isType/}{ih}, and $\mathit{ih}$, which directly
correspond to the assumptions available for any proof \(\texttt{ByInd} ·P'\).
We then elaborate the case branches with type \genid{Type/}{ih} and witness
\genid{isType/}{ih}, producing the collection of elaborated case bodies
\(\{\absu{\lamLam}{\vars{a_i}}{t_i'}\}_{i=1..\#\Delta}\) (since witness
\genid{isType/}{ih} is a variable, it will elaborate to itself).

In the conclusion of \textsc{[Mu]}, to elaborate the entire μ-expression we use the
generic function \texttt{mu} of Figure \ref{fig:elab-interface}, instantiating it with
the datatype's elaborated (inductive) signature functor $D\splab{FI}$ and proof
$\mathit{pos}$ it is positive. We also give \texttt{mu} the elaborated scrutinee
$t'$ and motive $P'$. 
The final argument to \texttt{mu} is a proof of type \(\texttt{ByInd} ·P\).
Within the body of this λ-expression, we invoke $x.2$ with a lifted motive,
giving it the elaborated case branches extended by assumptions
$m$ and $\mathit{eq}$ introduced by lifting.
Under this context, the elaborated case bodies $t'_i$ are
expected to have a type produced by lifting $P{'}$, and (by a forward reference
to Theorem \ref{thm:elab-typecheck-sound}) they indeed have types convertible
with this expected type. The final arguments required for the lifted elimination
is some \(D\splab{FI} ·\genid{Type/}{ih}\) (given by $x$) and proof it is equal
to $x$ (proved by β).

Rule \textsc{[Mu']} is similar to \textsc{[Mu]}, so we describe only the
significant differences. Operator {\mup} is given a scrutinee $t$ of type $T$, and
expects a witness $\mathit{is}$ proving that \(\genid{Is/}{D} ·T\) for a
suitable datatype $D$; this is checked by using the auxilliary judgment for
elaborating case branches. In the conclusion, we elaborate the {\mup}-expression
using \texttt{mu'}, whose last argument must be a proof of type
\(\texttt{ByCases} ·P{'} ·T\ \mathit{is}\), similarly given by property lifting.


\paragraph{\textbf{Soundness Properties}}
The elaborations of terms (types) from the surface language have their
elaborated types (kinds) in the internal language:

\begin{theorem}[Type-preservation]
  \label{thm:elab-typecheck-sound} If\, \(\vdash \Gamma \elab \Gamma'\) then:
  \begin{itemize}
  \item If\, \(\Gamma \vdash K \elab K'\) then \(\Gamma\splab{G},\Gamma' \vdash
    K'\)
  \item If\, \(\Gamma \vdash T : K \elab T'\) then for some $K'$, 
    \(\Gamma \vdash K \elab K'\)
    and \(\Gamma\splab{G},\Gamma' \vdash T' : K'\)
  \item If\, \(\Gamma \vdash t : T \elab t'\) then for some $T'$, \(\Gamma \vdash
    T : ★ \elab T'\) and \(\Gamma \vdash t' : T'\)
  \end{itemize}
\end{theorem}

\subsection{Operational Semantics}
\label{ssec:mu-reduce}
\begin{figure*}
  \begin{subfigure}{1.0\linewidth}
    \[
      \begin{array}{c}
        \begin{array}{lll}
          |\ \mumat{is}{t}{P}{c_i\ \vars{a_i} ➔ t_i}_{i=1..n}\ |
          & = & \mumatu{|t|}{c_i\ |\vars{a_i}| ➔ |t_i|}_{i=1..n}
          \\ |\ \mufix{ih}{t}{P}{c_i\ \vars{a_i} ➔ t_i}_{i=1..n}\ |
          & = & \mufixu{ih}{|t|}{c_i |\vars{a_i}| ➔ |t_i|}_{i=1..n} 
        \end{array}
        \\ \\
        \begin{array}{cc}
          \infer{
          \mumatu{(c_j\ \vars{s})}{c_i\ \vars{a_i} ➔ t_i}_{i=1..n}
          \reduce \vars{[s/a_j]}t_j
          }{
          1 \le j \le n \quad \#\vars{s} = \#\vars{a_j}
          }
          &
            \infer{
            \mufixu{ih}{(c_j\ \vars{s})}{c_i\ \vars{a_i} ➔ t_i}_{i=1..n}
            \reduce \vars{[s/a_j]}[r/ih]t_j
            }{
            1 \le j \le n \quad \#\vars{s} = \#\vars{a_j}
            \quad r = \absu{λ}{x}{\mufixu{ih}{x}{c_i\ \vars{a_i}
            ➔ t_i}_{i=1..n}}
            }
        \end{array}
      \end{array}
    \]
  \end{subfigure}
  \caption{Erasure and reduction for μ and μ'}
  \label{fig:mu-op-sem}
\end{figure*}
Part of the unwieldiness of working directly with λ-encodings is their
size. Our datatype subsystem for Cedille addresses this by treating
datatypes and their constructors opaquely, giving μ- and {\mup}-expressions a
primitive operational semantics shown in Figure \ref{fig:mu-op-sem}.
Cedille's operational semantics is defined for untyped terms, i.e., for terms
after the erasure of annotations. To erase both μ- and {\mup}-expressions (also
Figure \ref{fig:mu-op-sem}) we erase
the scrutinee, the motive, any type or erased term arguments bound by
constructor patterns (indicated by $|\vars{a_i}|$), and use the erasures of
the branch bodies; in
μ'-expressions we also erase the witness $\mathit{is}$.

\begin{figure*}[h!]
  \begin{subfigure}{1.0\linewidth}
    \[
      \begin{array}{cc}
        \infer{
        \Gamma \vdash \mufixu{ih}{t}{c_i\ \vars{a_i} ➔ t_i}_{i=1..n} \elab |\texttt{mu}|\ t'
        (\absu{λ}{ih}{\absu{λ}{x}{x\ (\absu{λ}{\vars{a_i}}{t_i'})_{i=1..n}}})
        }{
        \Gamma \vdash t \elab t'
        \quad (\Gamma \vdash t_i \elab t_i')_{i=1..n}
        }
        &
          \infer{
          \Gamma \vdash \mumatu{t}{c_i\ \vars{a_i} ➔ t_i}_{i=1..n} \elab
          |\texttt{mu'}|\ t'\ (\absu{λ}{x}{x\ (\absu{λ}{\vars{a_i}}{t_i'})_{i=1..n}})
          }{
          \Gamma \vdash t \elab t'
          \quad (\Gamma \vdash t_i \elab t_i')_{i=1..n}
          }
      \end{array}
    \]
  \end{subfigure}
  \caption{\fbox{\(\Gamma \vdash t \elab t'\)} Elaboration of pure (post-erasure)
    terms (shown: μ and μ')}
  \label{fig:elab-realizer}
\end{figure*}

\paragraph{\textbf{Soundness Properties}}
To show that our extension of the operational semantics is sound with respect to
that of the target language, we must introduce an auxilliary
judgment for \textit{elaboration of pure (post-erasure) terms} whose rules
mirror those for elaborating annotated terms -- the rules for this judgment for μ and {\mup} are
listed in Figure \ref{fig:elab-realizer}.

\begin{theorem}[Value Preservation for μ and μ']
  \label{thm:value-preservation}

  The elaborations of μ- and \mup-expressions and the elaborations of
  the terms they single-step are joinable:
  \begin{itemize}
  \item If\, \(\Gamma \vdash \mufixu{\mathit{ih}}{(c_j\ \vars{s})}{c_i\ \vars{a}
      ➔ t_i}_{i=1..n} \elab e_1\), and \\
    \(\mufixu{ih}{(c_j\ \vars{s})}{c_i\ \vars{a} ➔ t_i}_{i=1..n} \reduce t\),
    and
    \(\Gamma \vdash t \elab e_2\), then there exists some $e_3$ such that \(e_2
    \reduce^* e_3\) and \(e_2 \reduce^* e_3\)
  \item If\, \(\Gamma \vdash \mumatu{(c_j\ \vars{s})}{c_i\ \vars{a_i} ➔
      t_i}_{i=1..n} \elab e_1\), and \\ \(\mumatu{(c_j\ \vars{s})}{c_i\ \vars{a_i} ➔ t_i}_{i=1..n}
    \reduce t\), and \(\Gamma \vdash t \elab e_2\), then there exists some $e_3$
    such that \(e_1 \reduce^* e_3\) and \(e_2 \reduce^* e_3\)
  \end{itemize}
\end{theorem}

Finally, Theorem \ref{thm:cbn-normalization} states the termination guarantee of
our datatype subsystem.
\begin{theorem}[Call-by-name Normalization]
  \label{thm:cbn-normalization} 
  \ \\If\, \(\Gamma \vdash t : D \elab t'\) and
  \(\indel{D}{R}{\Delta}{\Theta}{\mathcal{E}} \in \Gamma\), and if $t$ is a
  closed term, then $|t'|$ is call-by-name normalizing.
\end{theorem}
\section{Related Work}
\label{sec:related-future-work}
\paragraph{\textbf{λ-encodings in CDLE}}
This work builds upon \cite{FBS18_Efficient-Mendler,FDJS18_CoV-Ind} which
generically derives induction (and CoV induction) in CDLE for λ-encoded
datatypes arising as the least fixed point of a class of type schemes
generalizing covariant functors.
Our elaborator interface was derived from these
developments: we repackaged the facilities of the generic library to implement
CoV pattern matching in the surface language without revealing implementation details.

\paragraph{\textbf{TCBs in ITPs}}
Many interactive theorem provers (ITPs) have large trusted computing bases
(TCBs). For example, Coq's kernel is $\sim$30K OCaml LoC, and some provers like
Agda \cite{No17_Agda} ($\sim$100K Haskell LoC) have no kernel.
But, there is much interest in verifying provers
themselves~\cite{DM15_Milawa-Theorem-Prover-Sound,
  Ha06_Self-Verification-of-HOL} and thus practical interest in keeping their
kernels small \cite{App01_Foundational-Proof-Carrying-Code}.

\cite{DM12_Elab-Inductive-Definitions} shares with us this goal, describing the
elaboration a language with inductive definitions, pattern matching, and
recursion to a simpler core theory.
They show how to
translate datatype declaration to Martin-L{\"o}f type theory extended with a
universe of positive inductive types and description labels. In comparison, our
core theory has no inductive primitives and elaboration produces explicit
proofs of positivity rather than elaborating to types that are positive by
construction.

\cite{GMK06_Eliminating-Dependent-Pattern} show how \textit{dependent pattern
  matching} (\cite{Co92_Pattern-Match-DT}) can be elaborated to a use of a
datatype's dependent eliminator. CoV pattern matching in this paper is in many
respects less sophisticated than dependent pattern matching;
however, an interesting point of comparison is the treatment of CoV induction.
\cite{GMK06_Eliminating-Dependent-Pattern} accomplish this by providing
as the inductive hypothesis \(\texttt{Below}_{D}\ P\ x\), a large tuple
containing proofs that $P$ holds for all subdata of $x$. Functions analyzing a
static number of cases analyzed (e.g. \texttt{fib}) may easily make use of this,
but accessing a proof for dynamically computed subdata (e.g. the result of
\texttt{minusCoV} in \texttt{divide}) requires an inductive proof of a lemma
such as \(\texttt{Below}_{\texttt{Nat}}\ P\ (\texttt{suc}\ n) ➔ P\
(\texttt{minus}\ n\ m)\) (for any $m$), not required in our work (nor of
\cite{FDJS18_CoV-Ind}).


\paragraph{\textbf{Semantic Termination Checking}}
\cite{Ab10_Sized-Types} extends type theory with \textit{sized types}, allowing datatypes to be annotated
with size information and the type system guaranteeing that recursive calls
are made on arguments of decreasing size. Sized types require defining
alternative, size-indexed versions of datatypes and extension of the underlying
theory, whereas in Cedille every standard datatype declaration is defined with
the usual notation and automatically supports CoV induction. On the other hand,
sized types allow for even more powerful forms of recursive definitions. In
particular, the usual implementation of merge-sort, which is definable using
sized types, is not straight-forwardly expressible as CoV recursion as it
involves recursion on terms that are not subdata of the original list.

The Nax language, described by \cite{Ah14_Nax}, takes an approach to termination
checking similar to ours. In Nax, recursive functions are defined in terms of
Mendler-style recursion schemes, including CoV recursion and Mendler-style
induction, whereas in Cedille the μ-operator of Cedille provides users the ability to
write definitions using CoV \textit{induction}. On the other hand,
Nax soundly permits datatype definitions with \textit{negative} recursive
occurrences, possible because Nax restricts the \textit{usage} of negative
datatypes, whereas we opt for the more traditional approach
of restricting the rules for the \textit{formation} of datatypes.

\section{Conclusion and Future Work}
\label{sec:conclusion}
We have presented a datatype subsystem for Cedille that enjoys
both the expected conveniences (compact notation for datatype
declarations, case analysis, and fixpoint-style recursive definitions) and the
desirable feature of CoV induction derived of λ-encodings in CDLE.
We further showed that this subsystem does not require
extending CDLE by presenting inference rules for the additional language
constructs that elaborate to expressions in Cedille 1.0.0 (which lacks a
datatype subsystem), and showing important soundness properties of the
types and operational semantics of elaborated terms with respect to the surface
language.

One immediate usability concern is the proliferation of explicit type coercions
in the case branches of μ- and \mup-expressions. We already automatically infer
the necessary type coercions for constructor arguments in the 
\textit{expected type} of case branches using the subtyping judgment in Figure
\ref{fig:elab-positivity}; this can be further integrated into the type
system so that type coercions in the \textit{bodies} of case branches need not
be explicitly coerced by the programmer, either.

Another direction is extending our datatype subsystem to support
\textit{zero-cost reuse} for programs and data, derived generically in CDLE by
\cite{DFS18_GenZC-Reuse}. One modest step would be to extend definitional
equality in the surface language so that constructors of different datatypes are
considered equal when their elaborated λ-expressions are, allowing
users to derive reuse \textit{manually} for datatypes and functions. More
ambitiously, a higher level syntax (such as \textit{ornaments}
\cite{Mc10_Ornaments,DM14_Transport-Ornament}) would allow programmers to define
one type (like \texttt{Vec}) in terms of another (like \texttt{List}) by
describing the function or relation on terms of the latter to the indices of the
former. Such definitions could then be elaborated using generic
zero-cost reuse combinators.

\begin{acks}                            
  We gratefully acknowledge NSF
  support under award 1524519, and DoD support under award
  FA9550-16-1-0082 (MURI program).
\end{acks}

\bibliography{icfp19-bib}

\end{document}